\documentclass[12pt]{article}
\usepackage[margin=1in]{geometry}
\usepackage{amsmath}
\usepackage{setspace}
\usepackage{graphicx}
\usepackage{epstopdf}
\usepackage[hidelinks]{hyperref}
\usepackage{natbib}
\usepackage{tikz}
\usepackage{multirow}
\usepackage{booktabs}
\usepackage{authblk}

\title{The role of post intercurrent event data in the estimation of hypothetical estimands in clinical trials}

\newcommand{\etal}{\textit{et al }}
\newcommand{\Var}{\mbox{Var}}
\newcommand{\Cov}{\mbox{Cov}}
\newcommand\ci{\perp\!\!\!\perp}

\author[1]{Jonathan W. Bartlett}
\author[2]{Rhian M. Daniel}
\affil[1]{Department of Medical Statistics, London School of Hygiene \& Tropical Medicine, London, WC1E 7HT, UK}
\affil[2]{Division of Population Medicine, Cardiff University, Cardiff, UK}

\date{}

\begin{document}
\maketitle

\large

\abstract{Estimation of hypothetical estimands in clinical trials typically does not make use of data that may be collected after the intercurrent event (ICE). Some recent papers have shown that such data can be used for estimation of hypothetical estimands, and that statistical efficiency and power can be increased compared to using estimators that only use data before the ICE. In this paper we critically examine the efficiency and bias of estimators that do and do not exploit data collected after ICEs, in a simplified setting. We find that efficiency can only be improved by assuming certain covariate effects are common between patients who do and do not experience ICEs, and that even when such an assumption holds, gains in efficiency will typically be modest. We moreover argue that the assumptions needed to gain efficiency by using post-ICE outcomes will often not hold, such that estimators using post-ICE data may lead to biased estimates and invalid inferences. As such, we recommend that in general estimation of hypothetical estimands should be based on estimators that do not make use of post-ICE data.}

\section{Introduction}
The ICH E9 addendum on estimands and sensitivity analyses lays out a framework for choosing and defining the treatment effect estimand of interest in a clinical trial \citep{ICHE9Addendum}. It introduces the notion of intercurrent events (ICEs) --- events after initiation of treatment that affect the interpretation or existence of patient outcomes, and a series of `strategies' for handling ICEs in the definition of the estimand. Use of the so-called hypothetical strategy involves targeting an effect of the randomised treatment under some hypothetical scenario for the ICE, for example what would have happened in the trial had the ICE in question been somehow prevented from occurring.

Historically such hypothetical estimands have typically been estimated only using data collected on patients up until they experience the ICE being handled by the hypothetical strategy. Recently however it has been shown that such estimands can be estimated using methods from the causal inference literature that can exploit data collected after the ICE occurrence. Specifically, \cite{Olarte2023Hypothetical} showed that the G-formula can do this, while \cite{lasch2022estimators}, \cite{lasch2022simulation}, and \cite{lasch2024comparison} demonstrated that G-estimation can similarly use this data. In these papers, it was shown that under the modelling assumptions assumed, the hypothetical estimand can be estimated with greater precision than when only data up until the ICE is used.

In this paper we critically examine the potential for using data collected after ICEs for estimation of hypothetical estimands. In Section \ref{sec:estimators} we examine estimators for hypothetical estimands that do not and do exploit post-ICE data. In so doing we articulate the assumptions that need to be made in order for the use of post-ICE data in the analysis to result in an increase in statistical efficiency and we consider their plausibility in trials. We also establish certain equivalences between apparently different G-formula estimators and G-estimators. In Section \ref{sec:efficiency} we provide expressions for the repeated sampling variance of estimators that do and do not exploit post-ICE data, and hence derive expressions for how much precision and power can be improved by using post-ICE data. In Section \ref{sec:simulations} we report simulations empirically demonstrating these results, and in Section \ref{sec:conclusions} give some recommendations about whether and when trials should attempt to use post-ICE data in their statistical analysis.

\section{Estimators of hypothetical estimands}
\label{sec:estimators}
In this section we describe a series of estimators for hypothetical estimands, some that do not and some that do use post ICE data. To focus on the key issues, we consider a highly simplified setting in which the ICE occurs (or not) at a single post-baseline time point, the structure of which we describe first.

\subsection{Causal structure, estimand and identification}
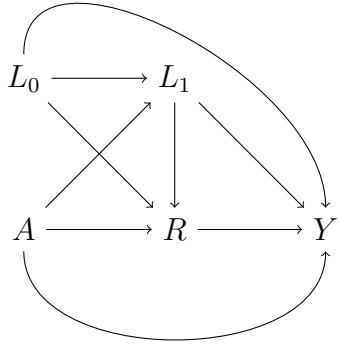
\begin{figure}
\begin{center}
    \begin{tikzpicture}
    \node (A) at (0.00,0.00) {$A$};
    \node (R) at (2.00,0.00) {$R$};
    \node (L_0) at (0.00,2.00) {$L_0$};
    \node (L_1) at (2.00,2.00) {$L_1$};
    \node (Y) at (4.00,0.00) {$Y$};
    
    \draw [->] (L_0) edge (L_1);
    \draw [->] (L_0) edge (R);
    \draw [->] (L_0) to[out=90,in=90] (Y);
    
    \draw [->] (A) edge (R);
    \draw [->] (A) to[out=-90,in=-90] (Y);
    \draw [->] (A) edge (L_1);

    \draw [->] (L_1) edge (Y);
    \draw [->] (L_1) edge (R);
    
    \draw [->] (R) edge (Y);    
    \end{tikzpicture} 
             \end{center}
    \caption{Directed acyclic graph (DAG) of a trial with baseline covariates $L_0$, randomised treatment $A$, post-baseline covariates $L_1$, intercurrent event indicator $R$, and outcome $Y$.}
    \label{fig:causalintroDAG_onetime}
\end{figure}

We let $A$ denote an individual patient's randomised treatment group. Let $R=1$ denote that the patient experiences the ICE and $R=0$ that they do not. The outcome is denoted $Y$, which we assume is continuous (we defer discussion of other outcome types to Section \ref{sec:conclusions}). $L_0$ and $L_1$ are baseline and post-baseline covariates that may affect the ICE $R$ (i.e.\ whether it occurs) and outcome $Y$.  Figure \ref{fig:causalintroDAG_onetime} shows a directed acyclic graph (DAG) that encodes the assumed causal structure between these variables.

The hypothetical estimand we focus on can now be defined in terms of potential outcomes \citep{lipkovich2020causal}. We let $Y^{a,r}$ denote the potential value of the outcome variable $Y$ if treatment is set to level $a$ ($a=0,1$) and the ICE to level $r$ ($r=0,1$). There is thus an implicit assumption that it could in principle be possible to set the level of the ICE $R$ for each patient. The effect of randomised treatment on mean outcome in the hypothetical scenario where we prevent the ICE is equal to $E(Y^{1,0})-E(Y^{0,0})$.

To identify and then estimate this hypothetical estimand, assumptions must be made \citep{Olarte2023Hypothetical}. The first is the so-called causal consistency assumption, which for the hypothetical estimand under no ICE defined above, states $Y=Y^{a,0}$ when $A=a$ and $R=0$. In words, this says that for a patient randomised in the actual trial to treatment group $a$ and who did not experience the ICE, their actual outcome $Y$ is identical to the potential outcome $Y^{a,0}$ they would have experienced in a hypothetical trial in which we prevent the ICE from occurring. The second is a conditional exchangeability assumption, that $R \ci Y^{a,r} | A=a, L_0, L_1$. This holds under the DAG shown in Figure \ref{fig:causalintroDAG_onetime}. The last is a positivity assumption: $P(R=0|A=a,L_0=l_0,L_1=l_1)>0$ for all possible values $l_0,l_1$ that can occur under each of the treatments $a=0,1$. This assumption says that across every level of treatment and the covariates $L_0$ and $L_1$ that can co-occur, there is a non-zero proportion of patients who would not have the ICE. This would be violated if for example for certain values of $L_1$, patients always experienced the ICE. This could occur in practice, if for example the ICE $R$ were rescue treatment and patients were always given rescue treatment if $L_1$ exceeded some threshold.

\subsection{Estimation ignoring post-ICE data using missing data methods}
We first consider estimation of $E(Y^{1,0})-E(Y^{0,0})$ not utilizing data on $Y$ in those for whom the ICE occurred ($R=1$), which currently is the standard approach adopted for estimation of hypothetical estimands. One perspective on this approach is that the potential outcomes $Y^{A,0}$ are missing for those patients who did experience the ICE ($R=1$). Under the earlier stated identification assumptions, as shown by \cite{Olarte2023Hypothetical}, the missing values in $Y^{A,0}$ are missing at random given $A,L_0,L_1$, and so we can adopt likelihood based approaches or an inverse probability of missingness weighting approach. Since likelihood based approaches such as observed data likelihood and multiple imputation (MI) are most commonly adopted, we focus on these. Specifically, we consider imputation based estimators of $E(Y^{1,0})-E(Y^{0,0})$. Although multiple imputations are typically generated, with inference obtained using Rubin's rules, when interest lies in estimating means or in fitting linear models the resulting MI point estimator is equivalent  to replacing each missing value by its fitted conditional mean \citep{wolbers2022standard}. As such, in the following we consider estimation where each missing no-ICE outcome is replaced by its fitted conditional mean. Thus we assume that we specify a model
\begin{equation}
    E(Y|A,L_0,L_1,R=0)=g_1(A,L_0,L_1,\beta_1)
    \label{eq:ymodR0Aa}
\end{equation}
with parameter vector $\beta_1$. This could for example be the linear mean model
\begin{equation}
    E(Y|A,L_0,L_1,R=0) = g_1(A,L_0,L_1,\beta_1) = \beta_{1,0} + \beta_{1,A} A + \beta^{T}_{1,L_0} L_0 + \beta^{T}_{1,L_1} L_1
    \label{eq:g1LinearMod}
\end{equation}
We fit this model by ordinary least squares to those with $R=0$, yielding an estimate $\hat{\beta}_1$ of the vector of parameters $\beta_1=(\beta_{1,0},\beta_{A},\beta_{1,L_0},\beta_{1,L_1})$. Based on this fitted model, we replace the missing $Y^{A,0}$ values in those who experienced the ICE with their fitted conditional means
\begin{equation*}
g_1(A_i,L_{i0},L_{i1},\hat{\beta}_1)
\end{equation*}
The estimate of the hypothetical estimand effect is then given by the mean difference in outcomes between randomised arms, using the imputed values in those with $R=1$ and the actual outcomes in those with $R=0$:
\begin{align}
\hat{\Delta}_{\text{imp,unadj}} &= \frac{1}{n_1} \sum_{i: A_i=1} \left\{(1-R_i) Y_i + R_i g_1(1,L_{i0},L_{i1},\hat{\beta}_1)\right\} - \frac{1}{n_0} \sum_{i: A_i=0} \left\{(1-R_i) Y_i + R_i g_1(0,L_{i0},L_{i1},\hat{\beta}_1)\right\} 
\label{eq:delta_imp_unadj}
\end{align}
For the particular choice of model given in equation \eqref{eq:g1LinearMod}, this estimator is
\begin{align}
\hat{\Delta}_{\text{imp,unadj}} =& \frac{1}{n_1} \sum_{i: A_i=1} \left\{(1-R_i) Y_i + R_i (\hat{\beta}_{1,0} + \hat{\beta}_{1,A} + \hat{\beta}^{T}_{1,L_0} L_{i0} + \hat{\beta}^{T}_{1,L_1} L_{i1})\right\} \nonumber \\
& - \frac{1}{n_0} \sum_{i: A_i=0} \left\{(1-R_i) Y_i + R_i (\hat{\beta}_{1,0} + \hat{\beta}^{T}_{1,L_0} L_{i0} + \hat{\beta}^{T}_{1,L_1} L_{i1})\right\}
\label{eq:deltahat_imp_unadj}
\end{align}

The imputation estimator $\hat{\Delta}_{\text{imp,unadj}}$ does not exploit the fact that $L_0$ and $A$ are independent as a consequence of randomisation. To improve efficiency, trial analyses would typically adjust for the baseline covariates $L_0$. When $Y$ is continuous, the default approach is a linear regression model of the outcome on treatment $A$ and baseline covariates $L_0$. Thus a baseline covariate adjusted imputation estimator $\hat{\Delta}_{\text{imp,adj}}$ is the solution to the ordinary least squares estimating equations
\begin{equation*}
\sum^{n}_{i=1} \begin{pmatrix} 1 \\ A_i \\ L_{i0} \end{pmatrix} \left\{(1-R_i) Y_i + R_i g_1(A_i,L_{i0},L_{i1},\hat{\beta}_1) - (\gamma_0 + \Delta_{\text{imp,adj}} A_i + \gamma_{L0} L_{i0}) \right\} = 0
\end{equation*}
where again the actual outcome $Y_i$ is used in those who did not experience the ICE ($R=0$) and the imputed outcome $g_1(A_i,L_{i0},L_{i1},\hat{\beta}_1)$ is used in those for whom it did.

\subsection{Estimation ignoring post-ICE data using G-formula}
The hypothetical estimand can also be estimated without using post-ICE data using the G-formula method from causal inference \citep{Olarte2023Hypothetical}. A G-formula estimator of $E(Y^{1,0})-E(Y^{0,0})$ based on the model in equation \eqref{eq:ymodR0Aa} is
\begin{equation}
    \hat{\Delta}_{\text{gform,pre,unadj}} = \frac{1}{n_1} \sum_{i:A_i=1} g_1(1,L_{i0},L_{i1},\hat{\beta}_1) - \frac{1}{n_0} \sum_{i:A_i=0} g_1(0,L_{i0},L_{i1},\hat{\beta}_1)
    \label{eq:muhatpreGformL1Emp}
\end{equation}
This G-formula estimator predicts outcomes for each individual in each treatment arm under no ICE, using the empirical distribution of $(L_0,L_1)$ in each treatment arm. It exploits the fact that $A$ is randomly assigned, which means that the distribution of $(L_0,L_1)$ in each treatment group is representative of the $(L_0,L_1)$ distribution that would arise if the total population were assigned that treatment. For the particular choice of model given in equation \eqref{eq:g1LinearMod}, this G-formula estimator is 
\begin{align}
\hat{\Delta}_{\text{gform,pre,unadj}} =& \frac{1}{n_1} \sum_{i:A_i=1} \left(\hat{\beta}_{1,0} + \hat{\beta}_{1,A} + \hat{\beta}^{T}_{1,L_0} L_{i0} + \hat{\beta}^{T}_{1,L_1} L_{i1} \right)\nonumber \\
& - \frac{1}{n_0} \sum_{i:A_i=0} \left(\hat{\beta}_{1,0} + \hat{\beta}^{T}_{1,L_0} L_{i0} + \hat{\beta}^{T}_{1,L_1} L_{i1}\right) \nonumber \\
&= \hat{\beta}_{1,A} + \hat{\beta}^{T}_{1,L_0} \left\{\hat{E}(L_0|A=1) - \hat{E}(L_0|A=0)\right\} + \hat{\beta}^{T}_{1,L_1} \left\{ \hat{E}(L_1|A=1) - \hat{E}(L_1|A=0) \right\}
\label{eq:gform_pre_unadj_simple}
\end{align}
The first term $\hat{\beta}_{1,A}$ corresponds to the direct effect of $A$ on $Y$, while the third term corresponds to the effect mediated via $L_1$. The second term is in large samples close to zero, since randomisation ensures $E(L_0|A=1)=E(L_0|A=0)$.

Although it is not apparent from their corresponding expressions, $\hat{\Delta}_{\text{imp,unadj}}$ and $\hat{\Delta}_{\text{gform,pre,unadj}}$ are identical when the $g_1(.)$ model is a linear model that includes at least an intercept and main effect of $A$ (such as the one in equation \eqref{eq:g1LinearMod}). This is because in this case, the estimating equations that $\hat{\beta}_1$ solves include the two equations:
\begin{equation*}
\sum_{i: R_i=0} \begin{pmatrix} 1 \\ A_i \end{pmatrix} \left\{Y_i - g_1(A_i,L_{i0},L_{i1},\hat{\beta}_1) \right\} = 0
\end{equation*}
This means that
\begin{equation*}
\sum_{i:  A_i=1,R_i=0} \left\{Y_i - g_1(1,L_{i0},L_{i1},\hat{\beta}_1)\right\}  = 0
\end{equation*}
and 
\begin{equation*}
\sum_{i: R_i=0} \left\{Y_i - g_1(A_i,L_{i0},L_{i1},\hat{\beta}_1)\right\}  = 0
\end{equation*}
Together these imply
\begin{equation*}
\sum_{i: A_i=0, R_i=0} \left\{Y_i - g_1(0,L_{i0},L_{i1},\hat{\beta}_1)\right\}  = 0
\end{equation*}
These equalities mean we can express $\hat{\Delta}_{\text{imp,unadj}}$ as
\begin{align*}
\hat{\Delta}_{\text{imp,unadj}} &= \frac{1}{n_1} \sum_{i: A_i=1} \left\{(1-R_i) Y_i + R_i g_1(1,L_{i0},L_{i1},\hat{\beta}_1)\right\} - \frac{1}{n_0} \sum_{: A_i=0} \left\{(1-R_i) Y_i + R_i g_1(0,L_{i0},L_{i1},\hat{\beta}_1)\right\} \\
&= \frac{1}{n_1} \left\{ \sum_{\substack{i: A_i=1,\\ R_i=0}} Y_i + \sum_{\substack{i: A_i=1,\\R_i=1}} g_1(1,L_{i0},L_{i1},\hat{\beta}_1) \right\} - \frac{1}{n_0} \left\{ \sum_{\substack{i: A_i=0,\\R_i=0}} Y_i + \sum_{\substack{i:A_i=0,\\R_i=1}} g_1(0,L_{i0},L_{i1},\hat{\beta}_1) \right\} \\
&= \frac{1}{n_1}  \sum_{i: A_i=1} g_1(1,L_{i0},L_{i1},\hat{\beta}_1) - \frac{1}{n_0}  \sum_{i: A_i=0} g_1(0,L_{i0},L_{i1},\hat{\beta}_1) \\
&= \hat{\Delta}_{\text{gform,pre,unadj}}
\end{align*}

We can also consider an adjusted G-formula estimator that takes the predicted no-ICE outcomes $g_1(A_i,L_{i0},L_{i1},\hat{\beta}_1)$ and regresses these on treatment $A_i$ and the baseline covariates $L_{i0}$. The resulting treatment effect estimator, $\hat{\Delta}_{\text{gform,pre,adj}}$ solves the estimating equations
\begin{equation*}
\sum^{n}_{i=1} \begin{pmatrix} 1 \\ A_i \\ L_{i0} \end{pmatrix} \left\{g_1(A_i,L_{i0},L_{i1},\hat{\beta}_1) - (\gamma^*_0 +\Delta_{\text{gform,pre,adj}} A_i + \gamma^*_{L0} L_{i0}) \right\} = 0
\end{equation*}
Perhaps as we might expect given the earlier equivalence of $\hat{\Delta}_{\text{imp,unadj}}$ and $\hat{\Delta}_{\text{gform,pre,unadj}}$, the baseline adjusted estimators $\hat{\Delta}_{\text{imp,adj}}$ and $\hat{\Delta}_{\text{gform,pre,adj}}$ are identical when the $g_1(.)$ model is a linear model that includes at least an intercept and main effects of $A$ and $L_0$.  This is because in this case, the first halves of each of the sets of estimating equations, which are the only parts that appear to differ, are in fact the same, i.e.
\begin{equation}
\sum^{n}_{i=1} \begin{pmatrix} 1 \\ A_i \\ L_{i0} \end{pmatrix} \left\{(1-R_i) Y_i + R_i g_1(A_i,L_{i0},L_{i1},\hat{\beta}_1) \right\} = \sum^{n}_{i=1} \begin{pmatrix} 1 \\ A_i \\ L_{i0} \end{pmatrix} \left\{g_1(A_i,L_{i0},L_{i1},\hat{\beta}_1) \right\}
\label{eq:impgformesteqequal}
\end{equation}
As such, the two sets of estimating equations are the same and hence have the same solution. To show the preceding equality holds, we consider the third component, and note that the first and second components can be argued similarly. Since the $g_1(.)$ model is a linear model that includes a main effect of $L_0$, the least squares estimating equations solved by $\hat{\beta}_1$ include
\begin{equation*}
\sum_{R_i=0} L_{i0} \left\{ Y_i - g_1(A_i,L_{i0},L_{i1},\hat{\beta}_1)\right\} = 0
\end{equation*}
so that $\sum_{R_i=0} L_{i0} Y_i = \sum_{R_i=0} L_{i0} g_1(A_i,L_{i0},L_{i1},\hat{\beta}_1)$. Then the left hand side of the third component of equation \eqref{eq:impgformesteqequal} is equal to
\begin{align*}
& \sum^{n}_{i=1} L_{i0} \left\{ (1-R_i) Y_i + R_i g_1(A_i,L_{i0},L_{i1},\hat{\beta}_1) \right\}  \\
&= \sum_{R_i=0} L_{i0} Y_i + \sum_{R_i=1} L_{i0} g_1(A_i,L_{i0},L_{i1},\hat{\beta}_1) \\
&= \sum_{R_i=0} L_{i0} g_1(A_i,L_{i0},L_{i1},\hat{\beta}_1) + \sum_{R_i=1} L_{i0} g_1(A_i,L_{i0},L_{i1},\hat{\beta}_1) \\
&= \sum^{n}_{i=1} L_{i0} g_1(A_i,L_{i0},L_{i1},\hat{\beta}_1),
\end{align*}
as required. In summary, we see that when linear outcome models are used, imputation missing data estimators are identical to corresponding G-formula causal inference estimators that do not exploit post-ICE outcomes \citep{Olarte2023Hypothetical}.

\subsection{Estimation using post-ICE data with G-formula}
\label{sec:estimators-post-ice-gformula}
We now investigate causal inference estimators that make use of post-ICE outcomes, first considering G-formula. Compared to the G-formula estimators described previously, we now fit a model to outcomes $Y$ including those for whom the ICE occurred ($R=1$), with $R$ included in the model. That is we fit a model
\begin{equation}
    E(Y|A,L_0,L_1,R)=g_2(A,L_0,L_1,R,\beta_2)
    \label{eq:ymodg2}
\end{equation}
For example we might assume that
\begin{equation}
E(Y|A,L_0,L_1,R)= \beta_{2,0} + \beta_{2,A} A + \beta_{2,L_0} L_0 + \beta_{2,L_1} L_1 + \beta_{2,R} R
\label{eq:ymodg2simplestform}
\end{equation}
Having estimated $\beta_2$, a G-formula estimator of $E(Y^{1,0})-E(Y^{0,0})$ can then be constructed as
\begin{equation}
    \hat{\Delta}_{\text{gform,prepost,unadj}} = \frac{1}{n_1} \sum_{i:A_i=1} g_2(1,L_{i0},L_{i1},0,\hat{\beta}_2) - \frac{1}{n_0} \sum_{i:A_i=0} g_2(0,L_{i0},L_{i1},0,\hat{\beta}_2)
    \label{eq:muhatprepostGform}
\end{equation}
which contrasts the average predicted outcome under treatment under no ICE with the average predicted outcome under control without ICE. If for example we use the model given in equation \eqref{eq:ymodg2simplestform}, we have
\begin{align}
    \hat{\Delta}_{\text{gform,prepost,unadj}} &= \frac{1}{n_1} \sum_{i:A_i=1} \hat{\beta}_{2,0} + \hat{\beta}_{2,A} + \hat{\beta}_{2,L_0} L_{i0} + \hat{\beta}_{2,L_1} L_{i1} - \frac{1}{n_0} \sum_{i:A_i=0} \hat{\beta}_{2,0} + \hat{\beta}_{2,L_0} L_{i0} + \hat{\beta}_{2,L_1} L_{i1} \nonumber \\
    &= \hat{\beta}_{2,A} + \hat{\beta}_{2,L_0} \left(\hat{E}(L_0|A=1) - \hat{E}(L_0|A=0) \right) + \hat{\beta}_{2,L_1}  \left(\hat{E}(L_1|A=1) - \hat{E}(L_1|A=0) \right)
    \label{eq:gform-prepost-unadj-simple}
\end{align}

The estimator $\hat{\Delta}_{\text{gform,prepost,unadj}}$ differs from $\hat{\Delta}_{\text{gform,pre,unadj}}$ in general due to inclusion of those with $R=1$ in the estimation of $\beta_2$. Assuming both $g_1(.)$ and $g_2(.)$ models are correctly specified, we expect $\hat{\Delta}_{\text{gform,prepost,unadj}}$ to be more efficient than $\hat{\Delta}_{\text{gform,pre,unadj}}$. However, gains in efficiency can only be achieved by assuming that that some of the covariate effects (i.e.\ effects of $A$, $L_0$, $L_1$) are common between the no-ICE ($R=0$) and ICE ($R=1$) groups, or equivalently, that the effect of the ICE does not depend on these variables. This is because if separate models are fitted in these two groups, the fit from the $R=1$ plays no role in the predictions in equation \eqref{eq:muhatprepostGform}. If the $g_2(.)$ model is incorrectly specified, in general $\hat{\Delta}_{\text{gform,prepost,unadj}}$ will be a biased estimator.

We believe that models $g_2(.)$ that assume the effect of the ICE is the same irrespective of the values of (one or more of) $A$, $L_0$ and $L_1$ will often be incorrect. For example, it might often be plausible that the effect of $R$ on $Y$ will differ depending on $L_1$. Consider for example the case of a diabetes trial where $R$ represents receipt of insulin rescue treatment and $L_1$ denotes fasting plasma glucose at visit 1, and $Y$ denotes HbA1c at the second and final visit. Here we might expect larger effects of insulin rescue treatment $R$ on HbA1c $Y$ among those with higher FPG $L_1$ values as compared with those with lower FPG $L_1$ values. This is because those with lower FPG values are likely to have lower HbA1c values, and there is less scope for insulin rescue to reduce HbA1c when the starting value is not so high. Similarly, if the active treatment is superior to the control treatment, it may be the case that the benefit of active treatment is reduced if insulin rescue has been taken, implying an interaction between $A$ and $R$ in the model for $Y$.

Lastly, we note that as before we can also consider a G-formula estimator that exploits randomisation by regressing the predicted no-ICE outcomes on treatment $A$ and baseline covariates $L_0$. Thus a baseline adjusted G-formula estimator is given by $\hat{\Delta}_{\text{gform,prepost,adj}}$, which solves the ordinary least squares estimating equations
\begin{equation}
\sum^{n}_{i=1} \begin{pmatrix} 1 \\ A_i \\ L_{i0} \end{pmatrix} \left\{g_2(A_i,L_{i0},L_{i1},0,\hat{\beta}_2)  - (\gamma_0 + \Delta_{\text{gform,prepost,adj}} A_i + \gamma_{L0} L_{i0}) \right\} = 0
\label{eq:gform_prepost_adj_esteq}
\end{equation}
As before, this estimator only exploits post-ICE outcomes $Y$ if the model $g_2(.)$ assumes the effect of the ICE $R$ to not depend on one or more of $A$, $L_0$, $L_1$.

\subsection{Estimation using post-ICE data with G-estimation}
\label{sec:estimators-g-estimation}
We now consider G-estimation utilising post-ICE outcomes as proposed by \cite{loh2020estimation} and further investigated by \cite{lasch2022estimators} and \cite{lasch2022simulation}. As described in the introduction, these papers demonstrated the potential for improved efficiency by using post-ICE outcomes relative to conventional estimators that do not exploit post-ICE data. We review these G-estimators from the perspective of structural nested direct effect models, as described by \cite{goetgeluk2008estimation}. In our setting, such a model assumes that
\begin{equation*}
E(Y^{ar} - Y^{0r}) = \phi(a,r;\upsilon)
\end{equation*}
for a specified function $\phi(a,r;\upsilon)$. The choice $\phi(a,r;\upsilon)=\upsilon a$ assumes the effect of randomised treatment is the same regardless of whether the ICE occurs or not, while the choice $\phi(a,r;\upsilon)=\upsilon_{0} a(1-r) + \upsilon_{1} ar$ allows the effect to differ depending on whether the ICE is fixed to zero or one. \cite{goetgeluk2008estimation} developed various estimators for $\upsilon$, including inverse probability weighted estimators based on specifying a model for the mediator, which here is the ICE $R$, those based on specifying a model for $E(Y|A,L_0,L_1,R)$, and doubly-robust estimators that involve specifying both of these models.

We consider first the structural nested direct effect model that assumes $E(Y^{ar} - Y^{0r})=\upsilon a$, and the so-called `unweighted G-estimation' approach that relies on specifying and fitting a model $g_2(A,L_0,L_1,R,\beta_2)$ for $E(Y|A,L_0,L_1,R)$. Having estimated $\beta_2$ by $\hat{\beta}_2$, a transformed or de-mediated outcome variable $\tilde{Y}_i$ is then calculated for each individual as
\begin{equation*}
\tilde{Y}_i = Y_i - \left\{g_2(A_i,L_{i0},L_{i1},R_i,\hat{\beta}_2) - g_2(A_i,L_{i0},L_{i1},0,\hat{\beta}_2)\right\}
\end{equation*}
Thus for individuals who did not experience the ICE ($R=0$), $\tilde{Y}=Y$, while for those who did ($R=1$), $\tilde{Y}$ adjusts the observed outcome to remove the effect of the ICE. The G-estimator of the hypothetical estimand is then given by the difference in means of $\tilde{Y}_i$ between treatment groups:
\begin{equation}
    \hat{\Delta}_{\text{gest,prepost,unadj}} = \frac{1}{n_1} \sum_{i:A_i=1} \tilde{Y}_i - \frac{1}{n_0} \sum_{i:A_i=0} \tilde{Y}_i
    \label{eq:gest_prepost}
\end{equation}
The G-estimator appears to more directly exploit the post-ICE outcomes than the G-formula estimator $\hat{\Delta}_{\text{gform,prepost,unadj}}$. In fact in certain situations they are identical. We can express the G-estimator as:
\begin{align*}
    \hat{\Delta}_{\text{gest,prepost,unadj}} =& \frac{1}{n_1} \sum_{i:A_i=1} \tilde{Y}_i - \frac{1}{n_0} \sum_{i:A_i=0} \tilde{Y}_i \\
    =& \frac{1}{n_1} \sum_{i:A_i=1} Y_i - \left\{g_2(1,L_{i0},L_{i1},R,\hat{\beta}_2) - g_2(1,L_{i0},L_{i1},0,\hat{\beta}_2)\right\} \\
    & - \frac{1}{n_0} \sum_{i:A_i=0} Y_i - \left\{g_2(0,L_{i0},L_{i1},R,\hat{\beta}_2) - g_2(0,L_{i0},L_{i1},0,\hat{\beta}_2)\right\}\\
    =& \frac{1}{n_1} \sum_{i:A_i=1} g_2(1,L_{i0},L_{i1},0,\hat{\beta}_2) - \frac{1}{n_0} \sum_{i:A_i=0} g_2(0,L_{i0},L_{i1},0,\hat{\beta}_2) \\
    & + \frac{1}{n_1} \sum_{i:A_i=1} Y_i - g_2(1,L_{i0},L_{i1},R,\hat{\beta}_2) - \frac{1}{n_0} \sum_{i:A_i=0} Y_i - g_2(0,L_{i0},L_{i1},R,\hat{\beta}_2) \\
    =& \hat{\Delta}_{\text{gform,prepost,unadj}}  \\
    & + \frac{1}{n_1} \sum_{i:A_i=1} Y_i - g_2(1,L_{i0},L_{i1},R,\hat{\beta}_2) - \frac{1}{n_0} \sum_{i:A_i=0} Y_i - g_2(0,L_{i0},L_{i1},R,\hat{\beta}_2)
\end{align*}
Reasoning as before, the two additional terms are zero, and thus $\hat{\Delta}_{\text{gest,prepost,unadj}}=\hat{\Delta}_{\text{gform,prepost,unadj}}$, when the $g_2(.)$ model is a linear model that includes at least an intercept and a main effect of $A$, which would almost always be the case. An example of this is the model given in equation \eqref{eq:ymodg2simplestform}. Equivalences between certain G-estimation and G-formula estimators have been previously shown by \cite{vansteelandt2009estimating}. Since the G-formula estimator $\hat{\Delta}_{\text{gform,prepost,unadj}}$ and G-estimator $\hat{\Delta}_{\text{gest,prepost,unadj}}$ are in these situations identical, any efficiency gains they offer relative to estimators that do not exploit post-ICE data are then of course identical as well. Moreover, our earlier conclusion that the G-formula estimator $\hat{\Delta}_{\text{gform,prepost,unadj}}$ only gains efficiency if the effect of the ICE $R$ on the outcome $Y$ is assumed to not depend on one or more of $A$, $L_0$ and $L_1$ also holds for $\hat{\Delta}_{\text{gest,prepost,unadj}}$. Indeed it is only if such an assumption is made that the G-estimator is actually using the post-ICE $Y$ outcomes (assuming model $g_2(.)$ includes an intercept and main effect of $A$).

To further improve efficiency, we might consider at the final stage of the G-estimator regressing $\tilde{Y}$ on $A$ and $L_0$, yielding an adjusted treatment effect estimator $\hat{\Delta}_{\text{gest,prepost,adj}}$ \citep{lasch2022estimators}. This estimator solves the estimating equations
\begin{equation*}
\sum^{n}_{i=1} \begin{pmatrix} 1 \\ A_i \\ L_{i0} \end{pmatrix} \left\{\tilde{Y}_i  - (\gamma_0 + \Delta_{\text{gest,prepost,adj}} A_i + \gamma_{L0} L_{i0}) \right\} = 0
\end{equation*}
As we might expect, the estimator $\hat{\Delta}_{\text{gest,prepost,adj}}$ is identical to the corresponding G-formula estimator $\hat{\Delta}_{\text{gform,prepost,adj}}$ under certain conditions. This is because the corresponding estimating equations only differ in the first part involving the outcome variable, and for this part, we have
\begin{align*}
\sum^{n}_{i=1} \begin{pmatrix} 1 \\ A_i \\ L_{i0} \end{pmatrix} \tilde{Y}_i &= \sum^{n}_{i=1} \begin{pmatrix} 1 \\ A_i \\ L_{i0} \end{pmatrix}  \left[ Y_i - \left\{g_2(A_i,L_{i0},L_{i1},R_i,\hat{\beta}_2) - g_2(A_i,L_{i0},L_{i1},0,\hat{\beta}_2)\right\} \right] \\
&= \sum^{n}_{i=1} \begin{pmatrix} 1 \\ A_i \\ L_{i0} \end{pmatrix} g_2(A_i,L_{i0},L_{i1},0,\hat{\beta}_2) + \sum^{n}_{i=1} \begin{pmatrix} 1 \\ A_i \\ L_{i0} \end{pmatrix} \left[ Y_i - g_2(A_i,L_{i0},L_{i1},R_i,\hat{\beta}_2) \right]
\end{align*}
The first part of the preceding expression is identical to the corresponding part in equation \eqref{eq:gform_prepost_adj_esteq} and the second part of the preceding expression equals zero so long as $g_2(.)$ is a linear model that at least includes an intercept and main effects of $A$ and $L_0$. Under such conditions, following the argument given in Section \ref{sec:estimators-post-ice-gformula}, the baseline covariate adjusted G-estimator $\hat{\Delta}_{\text{gest,prepost,adj}}$ only uses the post-ICE outcomes $Y$ when the $g_2(.)$ model assumes the effect of the ICE $R$ does not depend on one or more of $A$, $L_0$, and $L_1$.

Compared to the sequential G-estimation approach described above, the G-estimation approach proposed by \cite{loh2020estimation} involves fitting an additional model
\begin{equation*}
P(R=1|A,L_0,L_1) = h(A,L_0,L_1,\alpha)
\end{equation*}
with parameter vector $\alpha$. This could for example be a logistic regression model. An extended model is then fitted for $Y$ including $P=h(A,L_0,L_1,\hat{\alpha})$ as an additional covariate, for example:
\begin{equation*}
E(Y|A,L_0,L_1,R,P) = \beta_{3,0} + \beta_{3,A} A + \beta_{3,L_0} L_0 + \beta_{3,L_1} L_1 + \beta_{3,R} R + \beta_{3,P} P.
\end{equation*}
Then as before the outcomes $Y$ are adjusted to remove the estimated effect of the ICE $R$ using $\hat{\beta}_{3,R}$, and then the difference in means of these adjusted outcomes between randomised groups is calculated. \cite{loh2020estimation} stated that, provided the model for $P(R=1|A,L_0,L_1)$ is correct, the estimator of $\beta_{3,R}$ is unbiased even if the model for $E(Y|A,L_0,L_1,R,P)$ is misspecified. This may suggest that the resulting estimator of the hypothetical estimand is (asymptotically) unbiased provided only that $P(R=1|A,L_0,L_1)$ is correct. However, this G-estimator, and the one described previously, still rely on the assumption that $E(Y^{ar} - Y^{0r}) = \upsilon a$. This says that the effect of randomised treatment when the ICE $R$ is fixed at zero is the same as when it is fixed at one, or equivalently \citep{vansteelandt2009estimating}, that the effect of the ICE is the same under assignment to control as it is under assignment to active treatment. The assumption is a strong one, and again we envisage that often it will not hold. In the diabetes example, the effect of randomised treatment (active versus control) would typically be larger if insulin rescue was withheld ($E(Y^{1,0}-Y^{0,0})$) compared to if everyone was given insulin rescue ($E(Y^{1,1}-Y^{0,1})$).

Given the preceding concerns, one may therefore consider a more general direct effect model that specifies $E(Y^{ar} - Y^{0r})=\upsilon_{0} a(1-r) + \upsilon_{1} ar$. This allows the effect of randomised treatment to be different depending on whether the ICE is fixed to zero or one. Under this model, the parameter $\upsilon_{0}=E(Y^{1,0}-Y^{0,0})$ corresponds to our estimand of interest. G-estimators of $\upsilon_{0}$ can then be constructed, but using the theory developed by \cite{goetgeluk2008estimation} we show in Appendix \ref{app:gest-saturated-model} that under this more general structural nested direct effect model, post-ICE outcomes $Y$ play no role in inverse probability weighted G-estimators. For the unweighted and doubly-robust estimators, the post-ICE outcomes $Y$ only play a role if they are involved in estimation of the no-ICE conditional means $E(Y|A,L_1,R=0)$.

\section{Efficiency gain through using post-ICE data}
\label{sec:efficiency}
Simulations have demonstrated that estimators that exploit post-ICE data (e.g.\ G-formula and G-estimation) can be more efficient than those that do not \citep{Olarte2023Hypothetical,lasch2022estimators}. In Appendix \ref{app:asymptoticDist} we derive expressions for the asymptotic variance of the G-formula/G-estimation estimators in the simple case with no baseline covariates $L_0$ and a single variable $L_1$, under the assumption that
\begin{equation*}
E(Y|A,L_0,L_1,R)=g_2(A,L_0,L_1,R,\beta_2)=\beta_{2,0} + \beta_{2,A} A + \beta_{2,L_1} L_1 + \beta_{2,R} R
\end{equation*}
and that $\Var(Y|A,L,L_1,R)=\sigma^2$. We show that if $\Var(A,L_1|R=0)=\Var(A,L_1|R=1)$, the estimator $\hat{\Delta}_{\text{gform,pre,unadj}}$ that does not exploit post-ICE data is asymptotically normal with variance
\begin{align*}
\frac{\sigma^2 \kappa}{P(R=0)} + 2 \beta^2_{2,L_1} \left\{ \Var(L_1|A=0) +\Var(L_1|A=1) \right\} 
\end{align*}
whereas the estimator $\hat{\Delta}_{\text{gform,prepost,unadj}}$ that does exploit post-ICE data has asymptotic variance
\begin{align*}
\sigma^2 \kappa+ 2 \beta^2_{2,L_1} \left\{ \Var(L_1|A=0) +\Var(L_1|A=1) \right\} 
\end{align*}
where 
\begin{equation*}
\kappa = \frac{\Var(L_1|R) + 2 \Cov(A,L_1|R) (\tau_1-\tau_0) + \Var(A|R) (\tau_1-\tau_0)^2 }{\Var(A|R)\Var(L_1|R)-\Cov(A,L_1|R)^2}
\end{equation*}
and $\tau_a=E(L_1|A=a)$. The estimator that exploits post-ICE data thus has lower asymptotic variance, and the gain is larger as $P(R=0)$, the overall proportion of individuals who are ICE free across both arms, decreases. The variance $\sigma^2$ in the first term in each asymptotic variance represents variation in $Y$ not explained by $A$, $L_1$ and $R$. The second term, which is the same in the two asymptotic variances, depends on how much of the variation in $Y$ is explained by $L_1$. The relative advantage for the estimator that uses post-ICE data is larger when the proportion of variance of $Y$ explained by $L_1$ is smaller. The biggest advantage is conferred when $\beta_{2,L_1}=0$, meaning that all of the effect of $L_1$ on $Y$ is mediated via $R$. In this case, the ratio of the asymptotic variances (pre+post ICE data estimator to pre only estimator) is $P(R=0)$. Suppose the statistical power of the test of the null hypothesis that the hypothetical estimand is zero using the pre-ICE data estimator is $p$. Then the power of the test based on the estimator that uses post-ICE data can be shown to equal
\begin{equation}
\Phi\left(-1.96+\frac{1.96+\Phi^{-1}(p)}{\sqrt{P(R=0)}} \right)
\end{equation}
where $\Phi(.)$ denotes the cumulative distribution function of the standard normal distribution. For example, suppose the trial was sized to give $p=0.8$, or 80\% power, based on the estimator that does not use post-ICE data. Then with an overall proportion of individuals ICE free of 0.7, meaning 30\% of individuals experience the ICE (across the two arms) the power would be increased to 91.8\%. If instead the overall proportion experiencing an ICE is 15\%, a value that is quite typical in our experience, the power increases to 86.0\%. Recall that these calculations are in the best case scenario for the post-ICE estimator, where $\beta_{2,L_1}=0$. When some of the effect of $L_1$ on $Y$ is not mediated by the ICE $R$, the gains would be smaller.


\section{Simulations}
\label{sec:simulations}
In this section we report results of two simulation studies run to verify the asymptotic variance results given in Section \ref{sec:efficiency} and our assertions in Section \ref{sec:estimators-g-estimation} regarding bias of estimators that exploit post-ICE data. We simulated datasets of size $n=500$ with $P(A=1)=1/2$. The ICE indicator was generated using $P(R=0|A=a)=\pi_a$. We performed simulations for scenarios with $\pi_0=0.4,0.5,0.6,0.7,0.8$ and $\pi_1=\pi_0+0.1$, meaning that 10\% fewer patients in the active arm experience the ICE compared to the control arm. For simplicity, as in Section \ref{sec:efficiency} we did not simulate a baseline covariate $L_0$. The post-baseline variable $L_1$ was simulated using $L_1|A,R \sim N(A+R,1)$. This means that $P(R=1|A,L_1)$ followed a logistic regression model. In the first set of simulations, the outcome $Y$ was generated from $Y|A,L_1,R \sim N(A+L_1+R,1)$, while in the second set $Y$ was generated from $Y|A,L_1,R \sim N(A+L_1+R+(1/2)L_1R,1)$, thereby including an interaction between $L_1$ and $R$. Under both data generating models, the true hypothetical estimand was $1+1=2$. 

For each simulated dataset we estimated the hypothetical estimand first using $\hat{\Delta}_{\text{imp,pre,unadj}}$ (equivalently $\hat{\Delta}_{\text{gform,pre,unadj}}$) as given in equations \eqref{eq:deltahat_imp_unadj} and \eqref{eq:gform_pre_unadj_simple} (which we showed earlier are equal). Next we estimated the effect exploiting post-ICE data using $\hat{\Delta}_{\text{gform,prepost,unadj}}$ (equivalently  $\hat{\Delta}_{\text{gest,prepost,unadj}}$) as given in equation \eqref{eq:gform-prepost-unadj-simple}. Lastly we used Loh \etal's G-estimator $\hat{\Delta}_{\text{Loh}}$, which includes $P$ as a covariate, using a logistic regression model for $P(R=1|A,L_1)$, noting that this model is correctly specified under our data generating mechanism. We evaluated the estimators based on their bias and empirical standard errors. The latter was compared to the value calculated analytically using the asymptotic variance expressions derived in Appendix \ref{app:asymptoticDist}. Performance was evaluated using 10,000 simulations per scenario.

\begin{table}[ht]
\caption{Simulation results comparing imputation/G-formula estimator that does not use post-ICE data with G-formula/G-estimator that does, when the models for $Y$ used in these are correctly specified. $\pi_0$ and $\pi_1$ denote the proportions of individuals who do not experience an ICE. Emp. SE is the empirical standard error of the estimator and Asy. SE is the standard error calculated based on the derived asymptotic variance expressions.}
\centering
\begin{tabular}{rrrrrrrrrr}
\toprule
 & & \multicolumn{3}{c}{$\hat{\Delta}_{\text{imp,pre,unadj}}$ ($\hat{\Delta}_{\text{gform,pre,unadj}}$)} & \multicolumn{3}{c}{$\hat{\Delta}_{\text{gform,prepost,unadj}}$ ($\hat{\Delta}_{\text{gest,prepost,unadj}}$)} & \multicolumn{2}{c}{$\hat{\Delta}_{\text{Loh}}$}\\
 $\pi_0$ & $\pi_1$ & Bias & Emp. SE & Asy. SE & Bias & Emp. SE & Asy. SE & Bias & Emp. SE \\ 
\midrule
0.4 & 0.5 & -0.001 & 0.167 & 0.167 & 0.000 & 0.134 & 0.134 & 0.000 & 0.134 \\ 
  0.5 & 0.6 & 0.001 & 0.157 & 0.157 & 0.001 & 0.135 & 0.134 & 0.001 & 0.135 \\ 
  0.6 & 0.7 & 0.001 & 0.149 & 0.149 & -0.001 & 0.134 & 0.134 & -0.001 & 0.134 \\ 
  0.7 & 0.8 & 0.002 & 0.142 & 0.142 & 0.001 & 0.132 & 0.133 & 0.001 & 0.132 \\ 
  0.8 & 0.9 & 0.001 & 0.136 & 0.136 & 0.000 & 0.131 & 0.131 & 0.000 & 0.131 \\ 
\bottomrule
\label{sim-res-table-1}
\end{tabular}
\end{table}

Table \ref{sim-res-table-1} shows the results of the first set of simulations, for which all estimators are expected to be (asymptotically) unbiased because the models used for $Y$ were correctly specified. The bias results are inline with this, with empirical bias estimates very close to zero. The standard errors calculated using the asymptotic variance expressions we derived were very close to the empirical standard errors. As expected, the G-formula/G-estimator that used post-ICE data had lower empirical SE compared to the imputation/G-formula estimator that did not, and this difference reduced as the proportion of individuals remaining ICE free increased.

Lastly, Loh's G-estimator had empirical SE identical (to three decimal places) to the G-estimator that did not include $P$ as a covariate in the model for $Y$. If the model for $E(Y|A,L_0,L_1,R)$ (without $P$ as an additional covariate) is correctly specified, Theorem 6.2 of \cite{newey1994large} implies that the asymptotic variance of Loh's estimator is the same as that for the G-estimator $\hat{\Delta}_{\text{gest,prepost,unadj}}$, which does not use $P$ as a covariate in the model for $Y$. This is because consistency of Loh's estimator for the hypothetical estimand does not depend on $\alpha$ being estimated consistently. This in turn can be seen by the fact that if the model for $E(Y|A,L_0,L_1,R)$ is correctly specified, the true coefficient of $P$ in the model that additionally includes $P$ as a covariate is zero. As such, the estimators of the other parameters in $\beta_3$ remain consistent irrespective of whether $\alpha$ is consistently estimated. Consequently, when the model $E(Y|A,L_0,L_1,R)$ is correctly specified, inclusion of $P$ as a covariate in the model for $Y$ has no impact on the asymptotic variance.

\begin{table}[ht]
\caption{Simulation results comparing imputation / G-formula estimator that does not use post-ICE data with G-formula/G-estimator that does, when the model used by the estimator that uses post-ICE data is misspecified, by omitting an interaction between $L_1$ and $R$.}
\centering
\begin{tabular}{rrrrrrrrrr}
\toprule
 & & \multicolumn{3}{c}{$\hat{\Delta}_{\text{imp,pre,unadj}}$ / $\hat{\Delta}_{\text{gform,pre,unadj}}$} & \multicolumn{3}{c}{$\hat{\Delta}_{\text{gform,prepost,unadj}}$ / $\hat{\Delta}_{\text{gest,prepost,unadj}}$} & \multicolumn{2}{c}{$\hat{\Delta}_{\text{Loh}}$}\\
 $\pi_0$ & $\pi_1$ & Bias & Emp. SE & Asy. SE & Bias & Emp. SE & Asy. SE & Bias & Emp. SE \\ 
\midrule
0.4 & 0.5 & -0.001 & 0.168 & 0.167 & 0.246 & 0.158 & 0.134 & 0.247 & 0.158 \\ 
  0.5 & 0.6 & -0.001 & 0.157 & 0.157 & 0.201 & 0.153 & 0.134 & 0.200 & 0.153 \\ 
  0.6 & 0.7 & -0.001 & 0.148 & 0.149 & 0.155 & 0.148 & 0.134 & 0.153 & 0.148 \\ 
  0.7 & 0.8 & 0.001 & 0.142 & 0.142 & 0.109 & 0.144 & 0.133 & 0.106 & 0.144 \\ 
  0.8 & 0.9 & 0.001 & 0.136 & 0.136 & 0.060 & 0.138 & 0.131 & 0.057 & 0.138 \\ 
\bottomrule
\label{sim-res-table-2}
\end{tabular}
\end{table}

Table \ref{sim-res-table-2} shows the results of the second set of simulations, in which the true model generating $Y$ includes an $L_1R$ interaction, but the G-formula/G-estimator that uses post-ICE data does not include this interaction in its model for $Y$. As expected, the imputation/G-formula estimator that does not use post-ICE data continues to be unbiased, since the model used for $E(Y|A,L_1,R=0)$ remains correct. In contrast, the G-formula/G-estimator that does use the post-ICE data gives biased estimates, and, as one would expected, the bias is greater when the proportion of individuals experiencing an ICE is larger. We note that the SE calculated based on the asymptotic variance for the G-formula/G-estimator that uses post-ICE data is biased compared to the true empirical SE of the estimator, which is due to the fact that our asymptotic variance derivations assumed that the model used for $Y$ is correctly specified, which here it is not.

For Loh's G-estimator, we see that the empirical bias is essentially identical to that of the G-formula/G-estimator that uses post-ICE data. This is despite the fact the logistic model used for $P(R=1|A,L_1)$ was correctly specified. This is because, as described in Section \ref{sec:estimators-g-estimation}, this particular G-estimator still relies on assuming that $E(Y^{ar} - Y^{0r}) = \upsilon a$. This assumption is violated in our second simulation scenario by the presence of the $L_1 R$ interaction in the data generating model for $Y$. This is the case despite the lack of an $AR$ interaction in the data generating model. This is because the direct effect $E(Y^{ar} - Y^{0r})$ depends on both the direct effect of $A$ on $Y$ and the effect via the path $A \rightarrow L_1 \rightarrow Y$, and the latter effect differs depending on whether $R$ is set to zero or one when the data generating model for $Y$ includes the $L_1 R$ interaction.

\section{Conclusions}
\label{sec:conclusions}
For estimation of hypothetical estimands, estimators that make use of post-ICE outcome data can be more efficient than those that do not, increasing the statistical power to detect an effect. However, unless the proportions of patients experiencing the ICE are large, any gains in precision and power are likely to be modest in practice. Moreover, such gains in precision can only be achieved by assuming effects of randomised treatment and/or post-baseline confounders on the outcome are the same in those who do and do not experience the ICE. Recent papers demonstrating a power advantage for G-estimators that exploit post-ICE outcome data assumed we believe no effect modification for the effect of the ICE on outcome by randomised treatment or post-baseline confounders \citep{lasch2022estimators,lasch2022simulation}. We believe that in practice such effect modification will often exist, which would render these estimators biased. As such, we believe that in most cases it will be preferable to estimate hypothetical estimands using estimators that do not make use of post-ICE data and do not rely on assuming the absence of such effect modification.

We have only considered a highly simplified setting in which the ICE can occur (or not) at one time point and the final outcome is measured after this. As such, our conclusions may not generalise to the typical trial in practice where the ICE can occur at multiple time points and confounders and outcomes are measured repeatedly over time. However, we conjecture that our finding that efficiency can only be improved by assuming certain effects are common across those experiencing and not experiencing the ICE also applies in this more typical setting. This is because if one does not assume any such effects to be common, there is no information in the post-ICE data relevant to the prediction of patient outcomes had the ICE not occurred.

Although we have focused on the case of a continuous outcome $Y$ modelled using linear models, some of our conclusions apply more generally. Specifically, our conclusion that the G-formula estimator that uses post-ICE outcomes can only improve efficiency by assuming the effects of (some or all of) $A$, $L_0$ and $L_1$ on the outcome $Y$ are the same for those who do ($R=1$) and do not experience the ICE ($R=0$) applies to other outcome types. This is because only by assuming some of these effects are the same do the post-ICE outcome data play a role in estimation of the hypothetical estimand. For G-estimators, the conclusions of Appendix \ref{app:gest-saturated-model} regarding the role of post-ICE outcomes when the two controlled direct effects of randomised treatment are allowed to differ (i.e.\ the no-ICE effect $E(Y^{1,0})-E(Y^{0,0})$ is not assumed necessarily equal to the with-ICE effect $E(Y^{1,1})-E(Y^{0,1})$) similarly applies to other types of outcome $Y$.

The practical implication of our recommendation is that trial analyses targeting hypothetical estimands should adopt methods that only use data collected up until the ICE(s) handled by the hypothetical strategy. For continuous outcomes measured repeatedly over time, mixed models are commonly used, which rely on the missing at random assumption. Here we have instead focused on imputation estimators, since these can readily adjust for additional time-varying confounders of the effect of the ICE on outcome, which is generally necessary in order for MAR to be satisfied \citep{Olarte2023Hypothetical,olarte2025estimating}. Inverse probability of missingness weighted estimators can also be used. Such estimators in principle could be particularly attractive because unlike imputation based estimators, they rely on modelling the missingness process (here the occurrence of the ICE), which may be easier than specifying models for the repeated measures of outcome and time-varying confounders. Such inverse weighting estimators are generally less efficient than imputation estimators. Doubly-robust missing data estimators offer the potential for improved efficiency compared to inverse weighting estimators, but to the best of our knowledge have remained little used in trials \citep{bang2005doubly}.

\section*{Data}
R code for the simulation study is available at 
\newline \url{https://github.com/jwb133/hypothetical_post_ice_data}.

\section*{Funding}
This research was supported by UK Medical Research Council grants MR/T023953/1 and MR/T023953/2.

\section*{Conflict of interest}
JB’s past and present institutions have received consultancy fees for his advice on statistical methodology from AstraZeneca, Bayer, Novartis, and Roche. JB has in the past received consultancy fees from Bayer and Roche for statistical methodology advice.

\newpage

\appendix

\section*{Appendices}

\section{G-estimators of controlled direct effects}
\label{app:gest-saturated-model}
In this appendix we consider G-estimators as developed by \cite{goetgeluk2008estimation} under the (saturated) structural nested direct effect model
\begin{equation}
E(Y^{ar} - Y^{0r})=\upsilon_{0} a(1-r) + \upsilon_{1} ar
\label{app:eq-snde}
\end{equation}
\cite{goetgeluk2008estimation} considered a more general setting where the effect of treatment (in our setup, $A$) could be confounded by baseline variables $S$. In our randomised setting, there are no such variables because of randomisation, and so the variables denoted $S$ by \cite{goetgeluk2008estimation} are here empty. As a result, the functions denoted $q_k(S)$ by \cite{goetgeluk2008estimation} are zero. Following the recommendation for how to choose the functions labelled $d_k(X)$ by \cite{goetgeluk2008estimation}, we have for the model in equation \eqref{app:eq-snde} $d_r(A)=(A(1-R), AR)^T$. The IPW G-estimator based on equation 7 of \cite{goetgeluk2008estimation} then solves the estimating equation with estimating function
\begin{equation*}
P(R | A,L_1,\hat{\alpha})^{-1} \begin{pmatrix}A(1-R) - \hat{E}(A)(1-R) \\ A R - \hat{E}(A)R \end{pmatrix} \left\{Y -  \upsilon_{0} A(1-R) - \upsilon_{1} A R \right\} 
\end{equation*}
where $\hat{E}(A)$ is the sample proportion randomised to $A=1$ and $P(R|A,L_1,\hat{\alpha})$ denotes the predicted probability that the individual experienced an ICE (if $R=1$) or not (if $R=0$) based on a model fitted for $P(R=1|L_{1},A)$ with parameter $\alpha$. From inspection of the equation above we see that the first component of the estimating function involves $\upsilon_0$ but not $\upsilon_1$ and conversely the second component involves $\upsilon_1$ but not $\upsilon_0$. As such, only the first component equation is involved in estimation of the hypothetical estimand of interest $\upsilon_0$. All patients are involved in estimation of $\alpha$, but this does not involve the outcome $Y$, and the first component equation only involves $Y$ for those with $R=0$ due to the factor $(1-R)$ in the second term. As such, we conclude that the post-ICE outcomes (i.e.\ $Y$ from those with $R=1$) play no role in the IPW G-estimator of $\upsilon_0$, our parameter of interest.

Next, as described by \cite{goetgeluk2008estimation}, the unweighted G-estimator has estimating function given by their equation 19, setting the weight $W_i(\alpha)$ to one. In our notation, the estimating function under the model in equation \eqref{app:eq-snde} is thus
\begin{align*}
\begin{pmatrix}A(1-R) - \hat{E}(A)(1-R) \\ A R - \hat{E}(A)R \end{pmatrix} \left\{ Y - \hat{E}(Y|A,L_{1},R) \right\} \\
+ \hat{P}(R=0) \begin{pmatrix} A - \hat{E}(A)  \\ 0 \end{pmatrix} \left\{ \hat{E}(Y|A,L_1,R=0) - \upsilon_0 A \right\} \\
+ \hat{P}(R=1) \begin{pmatrix} 0 \\ A-\hat{E}(A)  \end{pmatrix} \left\{ \hat{E}(Y|A,L_1,R=1) - \upsilon_1 A \right\}
\end{align*}
Inspecting this estimating function, we again see that the first component of the estimating function involves only $\upsilon_0$ and the second only $\upsilon_1$. The outcomes $Y$ are only involved in the first term, and for the first component of this (which is involved in estimation for $\upsilon_0$), the term is zero for those with $R=1$. Thus we conclude that post-ICE outcomes ($Y$ for those with $R=1$) only play a role in estimation of $\upsilon_0$ if they are involved in estimation of $E(Y|A,L_1,R=0)$.

Lastly, we note that the above results also mean that doubly-robust estimators for $\upsilon_0$  similarly do not make use of post-ICE outcomes $Y$ under the model of equation \eqref{app:eq-snde}.

\section{Asymptotic distributions of estimators}
\label{app:asymptoticDist}
\subsection{Efficiency of G-formula/estimator that does not use post-ICE data}
We first consider the efficiency of the G-formula estimator $\hat{\Delta}_{\text{gform,pre,unadj}}$ that does not use the post-ICE outcomes. This is an example of a so-called two stage estimator \citep{newey1994large}, since we first estimate $\beta_1$ and then use $\hat{\beta}_1$ to estimate $E(Y^{1,0})-E(Y^{0,0})$ as given by equation \eqref{eq:muhatpreGformL1Emp}. We consider a setting with no $L_0$, a scalar $L_1$, and the following simple model for $E(Y|A,L_1,R=0)$
\begin{equation}
E(Y|A,L_1,R=0) = g_1(A,L_1,\beta_1) = \beta_{1,0} + \beta_{1,A} A + \beta_{1,L_1} L_1 = \beta^T_1 X
\label{eq:simplest_g1_model}
\end{equation}
where we let $X=(1,A,L_1)^T$. 

In order to use the theory from \cite{newey1994large}, we first express the estimator $\hat{\Delta}_{\text{gform,pre,unadj}}$ as the solution to the estimating equations
\begin{equation}
\sum^{n}_{i=1} \begin{pmatrix} 1 \\ A_i \end{pmatrix} \left\{ g_1(A_i,L_{i1},\hat{\beta}_1) - (\Delta_0 + \Delta_{\text{gform,pre,unadj}} A_i) \right\} = 0 . \label{eq:gform_pre_unadj_est_eq}
\end{equation}
Let $\Delta=(\Delta_0,\Delta_{\text{gform,pre,unadj}})^T$. Assuming that the model for $E(Y|A,L_1,R=0)$ is correct, then Theorem 6.1 of \cite{newey1994large} gives that $\sqrt{n} (\hat{\Delta}-\Delta)$ is asympotically normal with mean zero and variance covariance matrix
\begin{equation*}
G^{-1}_{\Delta} \Var\left\{ m_2(A,L_1,\Delta,\beta_1) + G_{\beta_1} \psi(Y,A,L_1,R,\beta_1)) \right\} G^{-1^{T}}_{\Delta}
\end{equation*}
where we will define and derive each of the new terms in this expression in turn. The term $m_2(A,L_1,\Delta,\beta_1)$ is the estimating function in the estimating equations \eqref{eq:gform_pre_unadj_est_eq}. That is,
\begin{align*}
m_2(A,L_1,\Delta,\beta_1) &= \begin{pmatrix} 1 \\ A \end{pmatrix} \left\{ g_1(A,L_{1},\beta_1) - (\Delta_0 + \Delta_{\text{gform,pre,unadj}} A) \right\} \\
&= \begin{pmatrix} 1 \\ A \end{pmatrix} \left\{ \beta_{1,0} + \beta_{1,A} A + \beta_{1,L_1} L_1 - (\Delta_0 + \Delta_{\text{gform,pre,unadj}} A) \right\}
\end{align*}
The term $G_{\Delta}$ is defined as
\begin{equation*}
G_{\Delta} = E\left[\frac{\partial}{\partial \Delta^{T}} m_2(A,L_1,\Delta,\beta_1) \right]
\end{equation*}
We have
\begin{align*}
\frac{\partial}{\partial \Delta^{T}} m_2(A,L_1,\Delta,\beta_1)  &= \frac{\partial}{\partial \Delta^{T}} \begin{pmatrix} 1 \\ A \end{pmatrix} \left\{ \beta_{1,0} + \beta_{1,A} A + \beta_{1,L_1} L_1 - (\Delta_0 + \Delta_{\text{gform,pre,unadj}} A) \right\} \\
&= \begin{pmatrix} -1 & -A \\ -A & -A \end{pmatrix}
\end{align*}
Since $E(A)=1/2$, we have
\begin{equation*}
G_{\Delta} = \begin{pmatrix} -1 & -1/2 \\ -1/2 & -1/2 \end{pmatrix}
\end{equation*}
and
\begin{equation*}
G^{-1}_{\Delta} = \begin{pmatrix} -2 & 2 \\ 2 & -4 \end{pmatrix}
\end{equation*}

The term $G_{\beta_1}$ is defined as
\begin{equation*}
G_{\beta_1} = E\left[\frac{\partial}{\partial \beta^{T}_1} m_2(A,L_1,\Delta,\beta_1) \right]
\end{equation*}
We have
\begin{align*}
\frac{\partial}{\partial \beta^{T}_1} m_2(A,L_1,\Delta,\beta_1)  &= \frac{\partial}{\partial \beta^{T}_1} \begin{pmatrix} 1 \\ A \end{pmatrix} \left\{ \beta_{1,0} + \beta_{1,A} A + \beta_{1,L_1} L_1 - (\Delta_0 + \Delta_{\text{gform,pre,unadj}} A) \right\} \\
&= \begin{pmatrix} 1 & A & L_1 \\
A & A & A L_1 \end{pmatrix}
\end{align*}
Thus $G_{\beta_1}$ is equal to
\begin{equation*}
G_{\beta_1} = \begin{pmatrix} 1 & 1/2 & E(L_1) \\
1/2 & 1/2 & (1/2) E(L_1|A=1) \end{pmatrix}
\end{equation*}
Let
\begin{equation*}
E(L_1|A) = \tau_0+(\tau_1-\tau_0) A
\end{equation*}
Then $G_{\beta_1}$ is given by
\begin{equation*}
G_{\beta_1} = \begin{pmatrix} 1 & 0.5 & 0.5 \tau_0 + 0.5 \tau_1 \\
0.5 & 0.5 & 0.5 \tau_1 \end{pmatrix}
\end{equation*}

The term $\psi(Y,A,L_1,R,\beta_1)$ is the so-called influence function of $\hat{\beta}_1$. It is given by
\begin{equation*}
\psi(Y,A,L_1,R,\beta_1) = - M^{-1} m_1(Y,A,L_1,R,\beta_1)
\end{equation*}
where $m_1(Y,A,L_1,R,\beta_1)$ is the estimating function corresponding to $\hat{\beta}_1$ and 
\begin{equation*}
M = E\left[\frac{\partial}{\partial \beta^T_1} m_1(Y,A,L_1,R,\beta_1) \right].
\end{equation*}
For the simple model in equation \eqref{eq:simplest_g1_model} that we are assuming, which is only fitted using those with $R=0$, we have
\begin{align*}
m_1(Y,A,L_1,R,\beta_1) &= \begin{pmatrix} 1 \\ A \\ L_1 \end{pmatrix} (1-R) \left\{Y - (\beta_{1,0} + \beta_{1,A} A + \beta_{1,L_1} L_1) \right\} \\
&= (1-R) X (Y-X^T \beta_1)
\end{align*}
where recall $X=(1,A,L_1)^T$. The derivative of this with respect to $\beta^T_1$ is then
\begin{align*}
\frac{\partial}{\partial \beta^T_1} m_1(Y,A,L_1,R,\beta_1) &= (R-1) X X^T
\end{align*}
Taking expectations of this, we have that
\begin{equation*}
M = - P(R=0) E[X X^T | R=0 ]
\end{equation*}
Thus the influence function of $\hat{\beta}_1$ is given by
\begin{equation*}
\psi(Y,A,L_1,R,\beta_1) = P(R=0)^{-1} E[X X^T | R=0 ]^{-1} m_1(Y,A,L_1,R,\beta_1)
\end{equation*}
Since $\psi(Y,A,L_1,R,\beta_1) = - M^{-1} m_1(Y,A,L_1,R,\beta_1)$, we have
\begin{align*}
\Var\left\{\psi(Y,A,L_1,R,\beta_1) \right\} &= M^{-1} \Var\left\{m_1(Y,A,L_1,R,\beta_1) \right\} M^{-1^{T}}
\end{align*}
Then by the law of total variance we have
\begin{align*}
\Var\left\{m_1(Y,A,L_1,R,\beta_1) \right\} &= \Var\left[ E\left\{m_1(Y,A,L_1,R,\beta_1) | A,L_1,R\right\} \right] + E\left[ \Var\left\{m_1(Y,A,L_1,R,\beta_1) | A,L_1,R \right\} \right] \\
&= E\left[ \Var\left\{m_1(Y,A,L_1,R,\beta_1) | A,L_1,R \right\} \right]
\end{align*}
using the fact that $E\left\{m_1(Y,A,L_1,R,\beta_1) | A,L_1,R\right\}=0$. Now assuming that $\Var(Y|A,L_1,R)=\sigma^2$ we have
\begin{align*}
\Var\left\{m_1(Y,A,L_1,R,\beta_1) | A,L_1,R \right\} &= \Var\left\{ (1-R)X(Y-X^T\beta_1) | A,L_1,R) \right\} \\
&= \Var\left\{ (1-R)XY | A,L_1,R) \right\} \\
&= (1-R) \sigma^2 XX^T
\end{align*}
and then
\begin{align*}
E\left[ \Var\left\{m_1(Y,A,L_1,R,\beta_1) | A,L_1,R \right\} \right] &= E\left[(1-R) \sigma^2 XX^T \right] \\
&= P(R=0) \sigma^2 E(XX^T|R=0)
\end{align*}
Then we have that the variance of the influence function is
\begin{align*}
\Var\left\{\psi(Y,A,L_1,R,\beta_1) \right\} &= M^{-1} \Var\left\{m_1(Y,A,L_1,R,\beta_1) \right\} M^{-1^{T}} \\
&= P(R=0)^{-1} E[ X X^T | R=0 ]^{-1} \\
& \times P(R=0) \sigma^2 E[X X^T | R=0] P(R=0)^{-1} E[ X X^T | R=0 ]^{-1} \\
&= \sigma^2 P(R=0)^{-1} E[ X X^T | R=0 ]^{-1}
\end{align*}

Now consider $\Var\left\{ m_2(A,L_1,\Delta,\beta_1) + G_{\beta_1} \psi(Y,A,L_1,R,\beta_1) \right\}$. We now argue that the covariance of the two terms in this sum is zero. This is because
\begin{align*}
&\Cov\left\{m_2(A,L_1,\Delta,\beta_1),\psi(Y,A,L_1,R,\beta_1) \right\} \\
=& E\left\{\Cov(m_2(A,L_1,\Delta,\beta_1),\psi(Y,A,L_1,R,\beta_1) | A,L_1,R) \right\} \\
&+ \Cov\left\{E(m_2(A,L_1,\Delta,\beta_1)|A,L_1,R), E(\psi(Y,A,L_1,R,\beta_1)|A,L_1,R) \right\}
\end{align*}
The covariance in the first term of the preceding equation is zero because $m_2(A,L_1,\Delta,\beta_1)$ is constant conditional on $A,L_1,R$. The second term is also zero, because $E(\psi(Y,A,L_1,R,\beta_1)|A,L_1,R)=0$. This means that 
\begin{align*}
& \Var\left\{ m_2(A,L_1,\Delta,\beta_1) + G_{\beta_1} \psi(Y,A,L_1,R,\beta_1) \right\} \\ &= \Var\left\{m_2(A,L_1,\Delta,\beta_1) \right\} + G_{\beta_1} \Var\left\{\psi(Y,A,L_1,R,\beta_1) \right\} G^{T}_{\beta_1}
\end{align*}
Thus the estimator has asymptotic variance
\begin{equation*}
G^{-1}_{\Delta} \Var\left\{ m_2(A,L_1,\Delta,\beta_1)\right\} G^{-1^{T}}_{\Delta} + G^{-1}_{\Delta} G_{\beta_1} \Var\left\{\psi(Y,A,L_1,R,\beta_1) \right\} G^{T}_{\beta_1} G^{-1^{T}}_{\Delta}
\end{equation*}

Further algebra shows that
\begin{equation*}
\Var\left\{m_2(A,L_1,\Delta,\beta_1) \right\} = \frac{1}{2} \beta^2_{1,L_1} \begin{pmatrix} \Var(L_1|A=0) + \Var(L_1|A=1) & \Var(L_1|A=1) \\ \Var(L_1|A=1) & \Var(L_1|A=1) \end{pmatrix}.
\end{equation*}
Then we have
\begin{align*}
G^{-1}_{\Delta} \Var\left\{ m_2(A,L_1,\Delta,\beta_1)\right\} G^{-1^{T}}_{\Delta} &= 2 \beta^2_{1,L_1} \begin{pmatrix} \Var(L_1|A=0) & -\Var(L_1|A=0) \\ -\Var(L_1|A=0) & \Var(L_1|A=0) + \Var(L_1|A=1) \end{pmatrix}
\end{align*}
The second diagonal element of the preceding quantity, which contributes to the asymptotic variance of $\hat{\Delta}_{\text{gform,pre,unadj}}$ is thus
\begin{equation*}
2 \beta^2_{1,L_1} \left\{ \Var(L_1|A=0) + \Var(L_1|A=1) \right\}
\end{equation*}

For the term $G^{-1}_{\Delta} G_{\beta_1} \Var\left\{\psi(Y,A,L_1,R,\beta_1) \right\} G^{T}_{\beta_1} G^{-1^{T}}_{\Delta}$, we have
\begin{align*}
G^{-1}_{\Delta} G_{\beta_1} &=  \begin{pmatrix} -2 & 2 \\ 2 & -4 \end{pmatrix}\begin{pmatrix} 1 & 0.5 & 0.5 \tau_0 + 0.5 \tau_1 \\
0.5 & 0.5 & 0.5 \tau_1 \end{pmatrix} \\
&= \begin{pmatrix}
-1 & 0 & -\tau_0 \\
0 & -1 & \tau_0 - \tau_1 
\end{pmatrix}
\end{align*}
From this it follows that the second diagonal element that contributes to the asymptotic variance of $\hat{\Delta}_{\text{gform,pre,unadj}}$, is given by
\begin{equation}
\sigma_{22} +2 \sigma_{23} (\tau_1 -\tau_0) + \sigma_{33} (\tau_1-\tau_0)^2
\label{eq:sigma22sigma23}
\end{equation}
where $\sigma_{22}$, $\sigma_{23}$ and $\sigma_{33}$ are the indicated elements in $\Var\left\{\psi(Y,A,L_1,R,\beta_1) \right\}$. The sub-matrix of $\Var\left\{\psi(Y,A,L_1,R,\beta_1) \right\}$ that corresponds to the effects of $A$ and $L$ can be shown to equal 
\begin{align*}
& \sigma^2 \left\{P(R=0) \Var((A,L) | R=0) \right\}^{-1} = \frac{\sigma^2}{P(R=0)} \frac{1}{\Var(A|R=0)\Var(L_1|R=0)-\Cov(A,L_1|R=0)^2} \\
& \times
\begin{pmatrix}
\Var(L_1|R=0) & - \Cov(A,L_1|R=0) \\
-\Cov(A,L_1|R=0) & \Var(A|R=0)
\end{pmatrix}
\end{align*}
Thus the quantity in equation \eqref{eq:sigma22sigma23} is equal to
\begin{align*}
\frac{\sigma^2  \left\{\Var(L_1|R=0) + 2 \Cov(A,L_1|R=0) (\tau_1-\tau_0) + \Var(A|R=0) (\tau_1-\tau_0)^2 \right\}}{P(R=0) \left\{\Var(A|R=0)\Var(L_1|R=0)-\Cov(A,L_1|R=0)^2\right\}}
\end{align*}

The asymptotic variance of $\hat{\Delta}_{\text{gform,pre,unadj}}$ is thus equal to
\begin{align}
\frac{\sigma^2 \left\{\Var(L_1|R=0) + 2 \Cov(A,L_1|R=0) (\tau_1-\tau_0) + \Var(A|R=0) (\tau_1-\tau_0)^2 \right\}}{P(R=0) \left\{\Var(A|R=0)\Var(L_1|R=0)-\Cov(A,L_1|R=0)^2\right\}} + \nonumber \\
2 \beta^2_{1,L_1} \left\{ \Var(L_1|A=0) +\Var(L_1|A=1) \right\} 
\label{eq:asympVarPreGForm}
\end{align}

\subsection{Estimator that uses post-ICE data}
Now consider the G-formula estimator $\hat{\Delta}_{\text{gform,prepost,unadj}}$ that uses post-ICE data. We assume that the following linear model is fitted:
\begin{equation*}
    E(Y|A,L_0,L_1,R)=g_2(A,L_0,L_1,R,\beta_2)=\beta_{2,0} + \beta_{2,A} A + \beta_{2,L_1} L_1 + \beta_{2,R} R.
\end{equation*}
Assuming the model is correctly specified, we have $\beta_{2,0}=\beta_{1,0}$, $\beta_{2,A}=\beta_{1,A}$, $\beta_{2,L_1}=\beta_{1,L_1}$. The estimator $\hat{\Delta}_{\text{gform,prepost,unadj}}$ then solves
\begin{equation}
\sum^{n}_{i=1} \begin{pmatrix} 1 \\ A_i \end{pmatrix} \left\{ g_2(A_i,L_{i1},0,\hat{\beta}_2) - (\Delta_0 + \Delta_{\text{gform,prepost,unadj}} A_i) \right\} = 0 . \label{eq:gform_prepost_unadj_est_eq}
\end{equation}
The factor $G_{\Delta}$ remains the same as in the case of the G-formula estimator that does not use post-ICE data. Moreover since $g_1(A_i,L_{i1},\beta_1)=g_2(A_i,L_{i1},0,\beta_2)$, the first contribution to the second factor in the asymptotic variance remains the same as in the estimator that only uses pre-ICE data. However, the second contribution in this factor changes due to the improved efficiency with which $\beta_1$ is estimated. We have
\begin{equation*}
G_{\beta_2} = E\left[\frac{\partial}{\partial \beta^{T}_2} m_2(A,L_1,\Delta,\beta_2) \right]
\end{equation*}
We have
\begin{align*}
\frac{\partial}{\partial \beta^{T}_2} m_2(A,L_1,\Delta,\beta_1)  &= \frac{\partial}{\partial \beta^{T}_2} \begin{pmatrix} 1 \\ A \end{pmatrix} \left\{ \beta_{1,0} + \beta_{1,A} A + \beta_{1,L_1} L_1 - (\Delta_0 + \Delta_{\text{gform,pre,unadj}} A) \right\} \\
&= \begin{pmatrix} 1 & A & L_1 & 0 \\
A & A & A L_1 & 0 \end{pmatrix}
\end{align*}
Thus $G_{\beta_2}$ is equal to
\begin{align*}
G_{\beta_2} &= \begin{pmatrix} 1 & 1/2 & E(L_1) & 0 \\
1/2 & 1/2 & (1/2) E(L_1|A=1) & 0 \end{pmatrix} \\
&= \begin{pmatrix} G_{\beta_1} & 0 \end{pmatrix}
\end{align*}
Hence 
\begin{equation*}
G^{-1}_{\Delta} G_{\beta_2} = \begin{pmatrix} G^{-1}_{\Delta} G_{\beta_1} & 0 \end{pmatrix}
\end{equation*}
Thus the only change to the variance of the estimator is that in the term
\begin{equation*}
\tilde{\sigma}_{22} +2 \tilde{\sigma}_{23} (\tau_1-\tau_0) + \tilde{\sigma}_{33} (\tau_1-\tau_0)^2
\end{equation*}
the covariance matrix terms correspond to $\Var\left\{\psi_2(Y,A,L_1,R,\beta_2) \right\}$, rather than of $\Var\left\{\psi_1(Y,A,L_1,R,\beta_1) \right\}$. The sub-matrix of $\Var\left\{\psi_2(Y,A,L_1,R,\beta_2) \right\}$ corresponding to $\beta_{2,A}$ and $\beta_{2,L}$ can be shown to equal
\begin{equation*}
\sigma^2 \left\{ P(R=0) \Var((A,L)^T | R=0) + P(R=1) \Var((A,L)^T | R=1) \right\}^{-1}
\end{equation*}
Thus the asymptotic variance of the estimator that exploits post-ICE data is
\begin{align}
\frac{\sigma^2 \left\{\overline{\Var}(L_1|R) + 2 \overline{\Cov}(A,L_1|R) (\tau_1-\tau_0) + \overline{\Var}(A|R) (\tau_1-\tau_0)^2 \right\}}{\overline{\Var}(A|R)\overline{\Var}(L_1|R)-\overline{\Cov}(A,L_1|R)^2} + \nonumber \\
2 \beta^2_{1,L_1} \left\{ \Var(L_1|A=0) +\Var(L_1|A=1) \right\} 
\label{eq:asympVarPrePostGForm}
\end{align}
where $\overline{\Var}(.|R)$ and $\overline{\Cov}(.,.|R)$ denote the corresponding elements of the average covariance matrix $P(R=0) \Var((A,L)^T | R=0) + P(R=1) \Var((A,L)^T | R=1)$.

If $\Var\left\{(A,L)^T | R=0\right\}=\Var\left\{(A,L)^T | R=1\right\}$ then this simply equals $\sigma^2 \Var\left\{(A,L)^T | R\right\}^{-1}$. In this case the estimator that uses post-ICE data has asymptotic variance
\begin{align}
\frac{\sigma^2 \left\{\Var(L_1|R) + 2 \Cov(A,L_1|R) (\tau_1-\tau_0) + \Var(A|R) (\tau_1-\tau_0)^2 \right\}}{\Var(A|R)\Var(L_1|R)-\Cov(A,L_1|R)^2} + \nonumber \\
2 \beta^2_{1,L_1} \left\{ \Var(L_1|A=0) +\Var(L_1|A=1) \right\} 
\label{eq:asympVarPrePostGFormEqCondVar}
\end{align}
which is identical to the asymptotic variance in equation \eqref{eq:asympVarPreGForm} except that in the first term, the factor $P(R=0)^{-1}$ does not appear. As such, the estimator that uses the post-ICE data has lower asymptotic variance, as expected.

\subsection{Simple parametric model}
To derive explicit expressions for the asymptotic variance we consider a particular parametric model. Suppose that $L_1|A,R \sim N(\lambda_0+\lambda_A A+\lambda_R R, \sigma^2_{L_1})$. Suppose further that $P(R=0|A=a)=\pi_a$. We now derive the quantities involved in the asymptotic variances (equation  \eqref{eq:asympVarPreGForm} and equation \eqref{eq:asympVarPrePostGForm}). First, we have
\begin{align*}
\Var(A|R=0)&=P(A=1|R=0)\left\{1-P(A=1|R=0)\right\} \\
&= \frac{P(A=1)P(R=0|A=1)}{P(R=0)} \left\{1-\frac{P(A=1)P(R=0|A=1)}{P(R=0)} \right\} \\
&= \frac{(1/2)\pi_1}{(1/2)\pi_0 + (1/2)\pi_1} \left\{1-\frac{(1/2)\pi_1}{(1/2)\pi_0 + (1/2)\pi_1} \right\} \\
&= \frac{\pi_1}{\pi_0+\pi_1} \left\{1-\frac{\pi_1}{\pi_0+\pi_1} \right\} \\
&= \frac{\pi_0 \pi_1}{(\pi_0+\pi_1)^2}
\end{align*}
and similarly one can show $\Var(A|R=1)=\frac{(1-\pi_0)(1-\pi_1)}{(2-\pi_0-\pi_1)^2}$. Next, we have
\begin{align*}
\Var(L_1|R=r)&= E(\Var(L_1|A,R=r) | R=r) + \Var(E(L_1|A,R=r) | R=r) \\
&= \sigma^2_{L_1} + \Var(\lambda_0+\lambda_A A + \lambda_R R|R=r) \\
&= \sigma^2_{L_1} + \lambda^2_A \Var(A |R=r) 
\end{align*}
and
\begin{align*}
\Cov(A,L_1|R=r) &= \Cov(A,\lambda_0+\lambda_A A + \lambda_R R|R=r) \\
&= \lambda_A \Var(A|R=r)
\end{align*}
We also have
\begin{align*}
E(L_1|A=1)-E(L_1|A=0) &= \lambda_0 + \lambda_A + \lambda_R E(R|A=1) - \left\{\lambda_0 + \lambda_R E(R|A=1) \right\} \\
&= \lambda_A + \lambda_R \left\{E(R|A=1)-E(R|A=0) \right\} \\
&= \lambda_A + \lambda_R (1-\pi_1-1+\pi_0) \\
&= \lambda_A + \lambda_R (\pi_0-\pi_1)
\end{align*}
and finally, we have
\begin{align*}
\Var(L_1|A=a) &= \Var(E(L_1|A=a,R)|A=a) + E(\Var(L_1|A=a,R)|A=a) \\
&= \Var(\lambda_0 + \lambda_A a + \lambda_R R | A=a) + \sigma^2_{L_1} \\
&= \lambda^2_R \Var(R | A=a) + \sigma^2_{L_1} \\
&= \lambda^2_R \pi_a (1-\pi_a) + \sigma^2_{L_1} 
\end{align*}

\end{document}